\documentstyle[12pt]{article}
\newcommand{\mrm}[1]{\mbox{\rm #1}}
\newcommand{\half}{{1\over 2}}
\newcommand{\bla}{\hspace{1cm}}
\newcommand{\beq}{\begin{equation}}
\newcommand{\eeq}[1]{\label{#1}\end{equation}}
\newcommand{\bea}{\begin{eqnarray}}
\newcommand{\eea}{\end{eqnarray}}

\newcommand{\Ref}[1]{eq.~(\ref{#1})}
\newcommand{\Frac}[2]{\frac{\displaystyle #1}{\displaystyle #2}}

\def\Tr#1{\mathop{\mrm Tr}\{ #1 \}}
\def\titlepage{\clearpage%
\setcounter{footnote}{0}%
\thispagestyle{empty}\pagestyle{plain}\pagenumbering{arabic}%
\kern1mm
\vskip15mm\normalsize}
\def\docnum#1{\hbox to \hsize{\hskip123mm\hbox{#1}\hss}}
\def\date#1{\hbox to \hsize{\hskip123mm\hbox{#1}\hss}}

\def\title#1{\vskip1em\begin{center}\Large\bf#1\end{center}\vskip2.5em}
\def\author#1{\vskip0.5em{\bf #1}\vskip0.5em}
\def\inst#1{\vskip0.3em{ #1}\vskip0.5em}
\def\abstract{\begin{center}{\bf Abstract}\end{center}\quotation}

\def\anotfoot#1#2{\vfill\noindent\underline{\hspace{6cm}}
\par\noindent #1) #2}
\begin{document}
\begin{titlepage}
\docnum{CERN--TH.6899/93}
\docnum{FTUV-23/93}
\vspace{2cm}
\title{QCD Matching Conditions at Thresholds}

\begin{center}
\author{Germ\'an Rodrigo}
\inst{Departament de F\'{\i}sica Te\`orica,\\
Universitat de Val\`encia, Val\`encia, Spain}
\inst{and}
\author{Arcadi Santamaria$^{*)}$}
\inst{TH-Division, CERN, 1211 Gen\`eve 23, Switzerland}
\end{center}
\vspace{2cm}

\begin{abstract}
   The use of MS-like renormalization schemes in QCD requires
an implementation of nontrivial matching conditions across
thresholds, a fact often overlooked in the literature. We
shortly review the use of these matching conditions in QCD
and check explicitly that the prediction
for $\alpha_s(M_Z)$, obtained by running the strong coupling
constant from the $M_\tau$ scale, does not substantially
depend on the exact value of the matching point chosen in
crossing the $b$-quark threshold when the appropriate matching
conditions are taken into account.
\end{abstract}

\vspace{2cm}
\vfill\noindent
CERN-TH.6899/93\\
May 1993
\anotfoot{*}{On leave of absence from Departament de
F\'{\i}sica Te\`orica,
Universitat de Val\`encia, and IFIC, Val\`encia, Spain.}
\end{titlepage}
\setcounter{page}{1}

     During the last years a great effort has been done at LEP
in order to measure the strong coupling constant $\alpha_s(M_Z)$
at the $Z$ mass scale \cite{L392,AL92,DE92,OP92}. This measurement
has been of crucial importance since
it allowed, within the experimental errors,
the running of the strong
coupling constant to be checked
from low energies to the electroweak scale.
However, by
going to higher orders in the renormalization group equations,
some confusion has arisen
in the literature on the different prescriptions one could
use to cross thresholds in the
evolution of the running coupling constant.
The problem appears when working in  $MS$-like renormalization
schemes: since these are mass-independent,
the decoupling theorem of Appelquist-Carazzone \cite{AC75}
is not fulfilled in ``non-physical''
quantities such as beta functions or coupling constants. Only
in physical quantities particles with large masses do decouple.
Logarithms of large masses
induced by the renormalization group
equations
in the couplings are cancelled against other logarithms that
appear in the calculation of physical observables. This is
obviously an inconvenient, since a lot of effort must be invested
in intermediate stages of a calculation to compute terms that
will cancel in physical quantities. To remedy this
problem the standard procedure has been the
use of the effective field theory language \cite{WI76,WE80,HA81}.
For example,
in QCD with a heavy quark and $N-1$ light quarks,
one builds a theory with $N$ quarks and an effective
field theory with $N-1$ quarks. Around the threshold of the heavy
quark one requires agreement of the two theories. This gives
a set of matching equations that  relate the couplings of
the theory with $N$ quarks with the couplings of the
theory with $N-1$ quarks. This way, below the heavy quark threshold
one
can work with the effective theory, but using effective couplings.
Then, by construction, decoupling is trivial.
This procedure is equivalent to other
renormalization schemes and allows us
to correctly obtain the asymptotic
value of the coupling constant. The price one has to pay is that
coupling constants might not be continuous at thresholds.
All this machinery is well established since the
early 80's \cite{WI76,WE80,HA81,twoloopmatch1,twoloopmatch2}
and matching conditions
were computed at the one-loop level \cite{WE80,HA81}
and at the two-loop level \cite{twoloopmatch1,twoloopmatch2}
for general gauge
theories. It also seems to be well known  for
people working with $GUTs$ \cite{HA81} where, in general,
special attention
has been paid to matching conditions at the different thresholds.
However, the fact that one has to use appropriate matching
conditions in passing thresholds has been frequently overlooked
in the running of the QCD coupling constant by just taking
a continuous coupling constant across thresholds. Then, the final
results depend strongly on the exact scale one uses to
connect the couplings \cite{ALt92,PE92,BC92,OP92}. To solve
this ambiguity some of the authors \cite{ALt92}
vary the matching scale
between $0.75$ and $2.5$ times the mass of the heavy
quark; others \cite{PE92,BC92} use directly $\mu_{th}=2m_q$,
and yet others \cite{OP92} determine $\alpha_s(M_Z)$ with
both $\mu_{th}=m_q$ and $\mu_{th}=2m_q$ and take the
average. Here we will show that when appropriate matching
conditions are taken into account the final answer does not
depend on the exact $\mu_{th}$ used to connect the couplings.

     Although most of the points discussed in this paper
are well known in some circles, given the confusion that exists
in the literature and the importance of the subject we found it
convenient to recall what the correct matching conditions are
and to show that when they are consistently taken into account
the dependence on the renormalization scale cancels (at least at the
order the calculation is done). Consistency
requires that if the evolution of the gauge coupling constant is done
at $n$ loops, matching conditions should be imposed using $n-1$ loop
formulae \cite{WE80}; then the residual dependence on the
renormalization scale is of order $n+1$. We will show this, explicitly,
when running the QCD gauge constant from the $\tau$ mass to the $Z$
mass passing through the $b$ threshold.


     The renormalization group equations in QCD for the strong
 gauge coupling constant and the quark masses are

 \bea
    \frac{d\alpha_s}{dt} &=&
    - \alpha_s^2\left( \beta_0 + \frac{\beta_1}{4\pi}
 \alpha_s + \frac{\beta_2}{(4\pi)^2} \alpha_s^2 + \cdots\right)
 \label{RGEgauge1}
 \\
    \frac{dm^2}{dt} &= & - 4\pi \left(\gamma_0 \frac{\alpha_s}{\pi} +
 \gamma_1 \left(\frac{\alpha_s}{\pi}\right)^2 + \cdots\right) m^2
 \label{RGEgauge2}
 \eea
      where
 \beq
    t = \frac{1}{4\pi} \log \left(\frac{\mu^2}{\mu^2_0}\right)
 \eeq{cc1}
 and $\mu_0$ is some reference point.

     The $\beta$ coefficients governing   the evolution of
 the gauge coupling constant are
 \bea
    \beta_0 &=& 11 - \frac{2}{3} N_F \nonumber\\
    \beta_1 &=& 102 - \frac{38}{3} N_F \\
    \beta_2 &=& \half \left( 2857 - \frac{5033}{9} N_F
    + \frac{325}{27} N_F^2 \right)
 \label{beta}
 \nonumber
 \eea
 with $N_F$ the number of quark flavours with mass lower
than the renormalization scale $\mu$. The first two coefficients are
scheme-independent (in $MS$-like schemes)
but the higher-order coefficients
depend on the renormalization conditions \cite{MA84}.
We give $\beta_2$ in the $\overline{MS}$ scheme.

The quark mass anomalous dimensions are
 \bea
    \gamma_0 &=& 2 \nonumber \\
    \gamma_1 &=& \Frac{101}{12} - \Frac{5}{18} N_F\ .
 \eea

Integration of \Ref{RGEgauge1} can be performed by
first inverting  the series on the right-hand side of \Ref{RGEgauge1}
and then integrating on $\alpha$ and $t$. Finally, one
can solve for $\alpha$, at the required order, by using iterative
methods. The  result we obtain can be written in the following
form
\beq
 \alpha_s(\mu) = \alpha^{(1)}_s(\mu)+\alpha^{(2)}_s(\mu)+
 \alpha^{(3)}_s(\mu)+\cdots
\eeq{alfamu}
where
$\alpha^{(1)}_s$, $\alpha^{(2)}_s$, $\alpha^{(3)}_s$ represent
the one-, two- and three-loop
contributions respectively, and are given by
 \beq
    \alpha^{(1)}_s(\mu) =
    \frac{\alpha_s(\mu_0)}{1 + \alpha_s(\mu_0) \beta_0 t}
 \eeq{UnLoop}

 \beq
    \alpha^{(2)}_s(\mu) = - (\alpha_s^{(1)})^2 b_1 \log K(\mu)
 \eeq{DosLoops}

 \beq
    \alpha^{(3)}_s(\mu) =
     (\alpha_s^{(1)})^3 \left(b_1^2 \log K(\mu)\left(\log K(\mu)-1\right)
     -(b_1^2-b_2)\left(1-K(\mu)\right)\right)\ ,
 \eeq{TresLoops}
 where
 \beq
 b_1= \frac{\beta_1}{4\pi \beta_0},\bla
 b_2 = \frac{\beta_2}{(4\pi)^2 \beta_0},\bla
 K(\mu) =  \frac{\alpha_s(\mu_0)}{\alpha_s^{(1)}(\mu)}
 \eeq{defs}
     The running quark mass can also be obtained analytically
at the one-loop level, from \Ref{RGEgauge2}. The result is
 \beq
    m^2(\mu) = m^2(\mu_0)
    (1+\alpha_s(\mu_0) \beta_0 t)^{-4\gamma_0/\beta_0}\ .
 \eeq{UnMassp}
 For $\alpha_s(\mu_0) \beta_0 t << 1$ the result is independent of
$\beta_0$ and can be written as  (we use $\gamma_0 = 2$)
 \beq
    m^2(\mu) = m^2(\mu_0) (1-8 \alpha_s(\mu_0)  t + \cdots )\ .
 \eeq{UnMass}
 This expression can be used as long as
$\mu$ is not very different from $\mu_0$; in particular we could
use it to simplify the matching conditions.

      Conventionally \cite{NA82,MA84},
      higher-order RGEs are solved by doing
 a power series expansion in $1/L$ with
 $L=\log(\mu^2/\Lambda^2)$. This solution is given
 in terms of the
 QCD scale $\Lambda$, which is defined in such a
way that it is renormalization-group-invariant
but scheme-dependent and the
 so-called invariant mass $\hat{m}$.
Passing of thresholds is implemented by
requiring continuity of the couplings at threshold, which in turn
requires defining different $\Lambda$'s for different $N_F$.
For our purposes we prefer to use the solutions given above
because they allow us to work more easily with
scale-dependent matching conditions.


     In the $MS$ scheme, or any of its
simple modifications such as $\overline{MS}$, the beta function
governing the running of the strong
coupling constant is independent of quark
masses. Then, contrary to what happens in
momentum-subtraction schemes ($MO$),
the Appelquist-Carazzone theorem \cite{AC75}
that states, when it can be applied,
that the heavy particles
decouple at each order of perturbation theory is not
realized in a trivial way. The decoupling of the heavy particles
is fulfilled in  physical quantities, but
coupling constants and  beta functions do not exhibit it.

     To obtain decoupling in $MS$ schemes we need to build
in the decoupling region, $\mu \ll M$, an
effective field theory that behaves as if only
the light degrees of freedom were present.  Matching
conditions  connect the parameters of the low-energy
effective Lagrangian
with the parameters of the full theory. This can be done
by evaluating some Green functions in perturbation theory
with both the full and the effective theories, then require they
are the same, up to terms $O(1/M)$, for values of the renormalization
scale just around the threshold. Then, the coupling
constant of the effective theory can be expressed
as a power series expansion
in the coupling of the full theory with coefficients that
depend on $\log(M/\mu)$.
In order to obtain a good approximation using only
the first few terms in the perturbative
expansion, we have to evaluate
matching conditions in a region where $M/\mu \sim O(1)$.
However, the results of these calculations should not depend
on exactly which $\mu$ is chosen.

     One-loop matching conditions have been obtained
in \cite{WE80,HA81} for a general gauge theory. To obtain
matching conditions in QCD at the two-loop level several approaches have
been pursued. Ovrut and Schnitzer \cite{twoloopmatch1}
computed the gluon self energies
at the two-loop level with both the full and the
effective theories and then
required matching in the threshold region.
We will
follow a more direct approach, devised by Bernreuther and Wetzel
\cite{twoloopmatch2}.
Using the $MO$ scheme  as an intermediate
stage, these authors were
able to
relate the $\overline{MS}$  coupling constant
$\alpha_{\overline{MS}}(\mu)$, with $N_F$ quark flavours, with
the gauge coupling constant $\alpha_{\overline{MS}}^-(\mu)$
of the effective field theory with $N_F-1$ quark flavours
in which a heavy quark with $m_{\overline{MS}}$ mass has
been integrated out. This is because in momentum subtraction
schemes the decoupling theorem is also realized in the coupling
constants. The obtained relation has the following form:
 \beq
    \alpha_{\overline{MS}}^- = \alpha_{\overline{MS}}\
 \left( 1 + \sum_{k=1}^{\infty} \alpha_{\overline{MS}}^k C_k (x) \right)
 \eeq{alfamo3}
     with
 \beq
    x=\frac{1}{4\pi}\log(m_{\overline{MS}}^2/\mu^2)
 \eeq{cc2}
     In order to calculate the coefficients $C_k$ Bernreuther and
Wetzel impose the RGEs, \Ref{RGEgauge1} and \Ref{RGEgauge2}, on
$\alpha_{\overline{MS}}$, $\alpha_{\overline{MS}}^-$ and
$m_{\overline{MS}}$,
and obtain for the first two coefficients
a set of coupled first-order linear differential equations
depending only on the beta and gamma
functions of the full and the effective
theories. By solving them they found for a general
$SU(N)$ group the following result valid for the $\overline{MS}$
scheme\footnote{The solution of the two (for two loops)
differential equations depends on two scheme-dependent
arbitrary constants. To fix them
one has to perform a complete calculation in the scheme one is
interested in.}
 \bea
\label{conexion}
    C_1 &=& \frac{2}{3}
    \left(x + \frac{1}{8\pi}\frac{\partial}{\partial D}
 \Tr I \vert_{D=4}\right)  \\
    C_2 &=& [C_1(x)]^2 + \frac{1}{2\pi}
    \left(\frac{5}{3}C_2(G) - C_2(R)\right) x
 + \frac{1}{9\pi^2}C_2(G) \nonumber \\
 & & - \frac{17}{96\pi^2}C_2(R)
 + \frac{1}{32\pi^2}\left(\frac{5}{3}C_2(G) - C_2(R)\right)
 \frac{\partial}{\partial D}\Tr{I}\vert_{D=4} \nonumber
 \eea
 with $C_2(G)=N$ and $C_2(R)=\Frac{N^2-1}{2N}$
the Casimir operator eigenvalues of the
adjoint and fundamental representations, respectively, and
$D$ the space-time dimension.
A technical point
about the trace of the identity in the Dirac space, $\Tr{I}$,
should be discussed here. Strictly, in a general $D$-dimensional
space, $D$ even, the only irreducible representation of
 \beq
    \{ \gamma^{\mu},\gamma^{\nu} \} = 2g^{\mu \nu}
 \eeq{usual}
 has dimension $f(D)=2^{D/2}$. However, we can choose
$\Tr{I}=f(D)=4$, or any other smooth function with $f(4)=4$.
Different choices of $f(D)$ lead to different trivial modifications
of the $\overline{MS}$ renormalization scheme. However,
as can be seen in
\Ref{conexion}, different choices give quite different matching
conditions. Hence, in order to specify
completely the renormalization scheme within the $MS$-like schemes
one should also specify which convention has been used
for $\Tr{I}$.
Here we will use the usual convention among phenomenology papers,
i.e. $\Tr{I}=4$. Then for QCD we have the following
two-loop matching condition to connect the theory with
$N-1$ quarks with the theory with $N$ quarks at the $q$-quark
threshold \cite{twoloopmatch2}
 $$
    \alpha_{N-1} (\mu_{th})  =  \alpha_N (\mu_{th})  +
  \Frac{\alpha_N^2(\mu_{th})}{3\pi} \log\Frac{m_q(\mu_{th})}{\mu_{th}}
 $$
\beq
     + \Frac{\alpha_N^3(\mu_{th})}{9\pi^2}
 \left(\left(\log\Frac{m_q(\mu_{th})}{\mu_{th}}\right)^2
 + \Frac{33}{4}\log\Frac{m_q(\mu_{th})}{\mu_{th}} + \Frac{7}{8} \right)
 .
 \eeq{DosEmp}
 Here, $\mu_{th}$ is the value at which we require matching. As
 commented, this equation is valid for arbitrary values of
 $\mu_{th}$ as long as it is not far away from $m_q(m_q)$.
 Should we use instead $\Tr{I} = 2^{D/2}$, the logarithm
in the second term would be changed to $\log \sqrt{2} m_q/\mu_{th}$
changing completely the behaviour of the matching conditions.
For example, it is clear from the above equation that one can always
choose a $\mu_{th}$
in order to make the coupling continuous across thresholds.
Using only the one-loop matching condition,  i.e.
only the first two terms
in the right-hand side of \Ref{DosEmp},
and with $\Tr{I}=4$, we should require
$\alpha_{N-1}(m_q) = \alpha_N(m_q)$.
Using the two-loop matching condition, the matching point is slightly
different \footnote{One can still impose
$\alpha_{N-1}(m_q)=\alpha_N(m_q)$
at the two-loop level, as Marciano does \cite{MA84}, but
this requires a slight modification of the $\overline{MS}$ scheme
in order to absorb the non-logarithmic term in \Ref{DosEmp}.}.
However, if a scheme with $\Tr{I}=2^{D/2}$ is used the
matching point is found around $\sqrt{2} m_q$. In what follows
we will keep the couplings discontinuous and check the invariance
of the final result with respect to the chosen matching scale $\mu_{th}$.

Equation (\ref{DosEmp}) can be simplified by using \Ref{UnMass} to
remove the dependence on the running mass and leave the result
in terms of $m_q \equiv m_q(m_q)$, the $\overline{MS}$
running mass evaluated
at its own value. Since we are running from low energies to high
energies it is also better to use the inverted equation. Thus,
our matching condition at the threshold of the quark $q$ will be:
 $$
    \alpha_N (\mu_{th})  =  \alpha_{N-1} (\mu_{th})  -
  \Frac{\alpha_{N-1}^2(\mu_{th})}{3\pi}
  \log\Frac{m_q}{\mu_{th}}
 $$
 \beq
     + \Frac{\alpha_{N-1}^3(\mu_{th})}{9\pi^2}
 \left(\left(\log\Frac{m_q}{\mu_{th}}\right)^2
 - \Frac{57}{4}\log\Frac{m_q}{\mu_{th}} - \Frac{7}{8} \right)\ .
 \eeq{invempalme}


     We start from a scale below the bottom-quark
threshold, where we know the value of the strong  coupling
constant\footnote{The value of $\alpha_3(M_\tau)$ has been
extracted  at the two-loop level from hadronic $\tau$ decays
in \cite{PI93}. We took their result directly.}
 ,  e.g. $M_{\tau}$
 \beq
 \begin{array}{l}
    M_{\tau} = 1776.9 \pm 0.7\ \mbox{\rm MeV} \\ \\
    \alpha_3(M_{\tau}) = 0.36 \pm 0.03\ ,
 \end{array}
 \eeq{cc4}
We use\footnote{
Our starting point for the quark masses are the so-called
Euclidean masses \cite{NA87}, $M_b^E=4.23\pm 0.05$~GeV and
$M_c^E =1.26 \pm 0.02$~GeV,
from which we extract the $\overline{MS}$ masses
$m_b\equiv m_b(m_b) = 4.3 \pm 0.2$~GeV  and
$m_c\equiv m_c(m_c) = 1.3 \pm 0.2$~GeV.}
\Ref{invempalme} with $\mu_{th}=M_\tau$ and $m_q=m_c$
to obtain $\alpha_4(M_\tau)$ in terms of $\alpha_3(M_\tau)$.
Then we
evolve $\alpha_4(\mu)$ until the $Z$ boson mass
scale by
imposing matching conditions at an arbitrary intermediate
scale $\mu_{th}$ around $m_b$. To run $\alpha_4(\mu)$ from
$M_\tau$ until $\mu_{th}$ we use \Ref{alfamu} with $\mu_0=M_\tau$ and
four-quark beta functions.
Then at $\mu_{th}$ we impose the matching condition
\Ref{invempalme} with $m_q=m_b$ to obtain $\alpha_5(\mu_{th})$
in terms of $\alpha_4(\mu_{th})$.
Finally, to run
 $\alpha_5(\mu)$ from $\mu_{th}$ to $M_Z$ we use again \Ref{alfamu},
 but now with $\mu_0=\mu_{th}$, and with five-quark beta functions.
The evolution is consistent, i.e. to the same order,
if $n$-loop beta functions are used together with matching conditions
evaluated at the $(n-1)$-loop level.
Firstly, we run $\alpha_s(\mu)$ at the one-loop order, \Ref{UnLoop},
with matching conditions at tree level, i.e.
 taking $\alpha_4(\mu_{th})=\alpha_5(\mu_{th})$ with
 $\mu_{th}$ around $m_b$.
 After that, we calculate $\alpha_s(M_Z)$ by
running $\alpha_s(\mu)$ with two-loop beta functions and
imposing matching conditions at the one-loop order,
\Ref{invempalme}, but
taking only the first two terms on its right-hand side.
And finally, we evaluate $\alpha_s(M_Z)$ according to
the three-loop evolution, \Ref{TresLoops}, with matching
conditions at two-loop level, \Ref{invempalme}.

 We show the final results in fig.~1.
 We can clearly see that, as expected, the variation of the final
prediction on $\alpha_5(M_Z)$, as we vary the matching point around
the bottom quark mass, is of the same order of magnitude as the
next-order corrections; for three-loop beta functions and
two-loop matching conditions it is practically flat. For
comparison purposes we also give the error bar induced from
the error in $\alpha_s(M_\tau)$.
Given the level of accuracy, two-loop beta
functions and one-loop matching conditions seem to be good enough for all
purposes.

In the preceding section we directly used the matching equation
to evaluate $\alpha_4(M_\tau)$ in terms of $\alpha_3(M_\tau)$. We
could proceed in that way because the mass of the $c$-quark and
the mass of the $\tau$ are not so different; the logarithms in
the matching equation are therefore not large. Alternatively one could
try to run $\alpha_3(M_\tau)$ until some intermediate scale
$\mu_{th}$ around the charm threshold. Then, impose \Ref{invempalme}
with $m_q=m_c$ to get $\alpha_4(\mu_{th})$ and run it until
the bottom-quark threshold. This time, since we are interested only
in the error induced by crossing the charm threshold we will
use \Ref{invempalme} with $m_q=m_b$ and $\mu_{th}=m_b$ fixed to
obtain $\alpha_5(m_b)$. Finally we run $\alpha_5(\mu)$ from
$m_b$ until $M_Z$. Of course this procedure should give, within
the level of precision of the order considered,
the same result
as before. In fig.~2 we give $\alpha_5(M_Z)$ as a function
of the matching point $\mu_{th}$ taken for the charm threshold.
Although now the result depends
on the matching scale $\mu_{th}$, this dependence is
always a next-order
correction as long as the matching conditions are implemented
correctly. Clearly this procedure is potentially very dangerous
since an incorrect use of matching conditions could lead to
a false strong dependence on the matching scale. A similar
consideration could be applied to the bottom quark threshold. Then,
probably the safest procedure to run $\alpha_3(M_\tau)$ until the
$Z$ mass would be to use first \Ref{invempalme} with
$m_q=m_c$ and $\mu_{th}=M_\tau$ to get $\alpha_4(M_\tau)$ in
terms of $\alpha_3(M_\tau)$, then use again \Ref{invempalme}
with
$m_q=m_b$ and $\mu_{th}=M_\tau$ to get $\alpha_5(M_\tau)$ in
terms of $\alpha_4(M_\tau)$. Finally we should run $\alpha_5(\mu)$
from $M_\tau$ until $M_Z$ with the full five-quark renormalization
group. This procedure is justified since the masses of the $b$-quark,
$c$-quark and $\tau$-lepton are not so different as to spoil the
validity of the matching equation. Working in this way we arrived at the
value\footnote{We include the error induced by the errors in the
quark masses, which is about $0.001$ in $\alpha_5(M_Z)$.}
 $\alpha_5(M_Z) = 0.123 \pm 0.004$,
in complete agreement with
our previous result.

      To conclude, we would like to remark on the following points:
\begin{itemize}
    \item{Only in $MO$-like schemes, where Appelquist-Carazzone
is realized in both beta functions and coupling constants,
the strong  coupling constant $\alpha_s(\mu)$
is continuous. In $\overline{MS}$-like schemes one should build
a low-energy effective field theory and write scale-dependent
matching conditions in order to connect the parameters of the
theories on both sides of the threshold $\mu_{th}$. Then, for general
values of $\mu_{th}$ the couplings are not continuous although
in the case of only one coupling constant it is always possible
to find a particular $\mu_{th}$ that makes the coupling continuous.}

    \item{Evolution is consistent,  i.e. to the same order, if the
evolution of the gauge coupling constant at the $n$-loop order
is accompanied by matching conditions at the $(n -1)$-loop level.}

    \item{Different choices for the trace in Dirac space,
i.e. $\Tr{I} = 4$ or
$\Tr{I} =  2^{D/2}$, give rise to different trivial
modifications of the $\overline{MS}$ scheme with quite different
matching conditions.
Should one insist on having
a continuous coupling across thresholds, it is clear from the
discussion that the precise matching point will depend
on the choice for the trace in Dirac space. For instance, working
with two-loop beta functions one should take
    \subitem{$\mu_{th} = m_b$\ \ if\ \ $\Tr{I}= 4$}
    \subitem{$\mu_{th} = \sqrt{2} m_b$\ \ if\ \ $\Tr{I} = 2^{D/2}$}
} \ .
\end{itemize}

     By running the strong coupling
constant from the $M_\tau$ scale to the $M_Z$ scale,
we have checked explicitly
that
the final answer is not sensitive to the exact value of the
matching point, $\mu_{th}$,
 used in crossing the $m_b$ threshold as long as
the right matching conditions are consistently
taken into account (fig.~1). Similar considerations apply
when crossing the $m_c$ threshold (fig.~2). Finally we have shown
that the correct result can be obtained by using the matching
conditions to find $\alpha_5(M_\tau)$ in terms of $\alpha_3(M_\tau)$
and then run it with the full five-quark renormalization group
until the $M_Z$ scale.

\section*{Acknowledgements}
We thank A. Pich for helpful discussions
on the subject of this paper and for a critical reading
of the manuscript. G. Rodrigo acknowledges the CERN theory group
for its hospitality during the preparation of this work
and the Conselleria de Cultura, Educaci\'o
i Ci\`encia de la Generalitat Valenciana for financial support.
This work has been supported in part by
CICYT, Spain, under grant AEN90-0040.

\vfil\eject
\section*{Figure captions}

{\bf Figure 1:} Strong coupling constant at the $M_Z$ scale, obtained
by running the coupling from its value at the
$M_\tau$ scale ($\alpha_3(M_\tau)=0.36\pm 0.03$), as a function of
the matching point taken to cross the $b$-quark threshold.
The
long-dashed line is obtained by using one-loop beta
functions and tree-level matching conditions. The dashed line is obtained
with two-loop beta functions and one-loop matching conditions, and the
solid line is obtained with three-loop beta functions and two-loop
matching conditions. Error bars on the final three-loop result
are given for comparison purposes with other $\alpha_s(M_Z)$ results.
\vskip 0.4cm\noindent
{\bf Figure 2:} Same as in fig.~1, but varying the matching point
around the $c$-quark threshold. The matching point for the
$b$-quark is now fixed at $m_b$.

\end{document}